


%

\magnification=1200
\overfullrule=0pt
\pageno=0

\rightline{McGill/92-32}
\rightline{hep-ph/9207266}
\rightline{(July, 1992)}
\bigskip\bigskip\bigskip
\centerline{\bf NAVIGATING AROUND THE ALGEBRAIC JUNGLE OF QCD:}
\medskip
\centerline{\bf EFFICIENT EVALUATION OF LOOP HELICITY AMPLITUDES}
\bigskip\bigskip\bigskip
\centerline{C.S. Lam$^*$}
\bigskip
\centerline{Department of Physics, McGill University, 3600 University
St.}
\medskip
\centerline{Montreal, P.Q., Canada H3A 2T8}
\bigskip\bigskip\bigskip\bigskip
\bigskip\bigskip\bigskip\bigskip
\bigskip\bigskip
\centerline{\bf Abstract}

\bigskip\bigskip\bigskip
A method is developed whereby spinor helicity techniques can be used
to
simplify the calculation of loop amplitudes. This is achieved by
using the Feynman-parameter representation where the offending
off-shell loop momenta do not appear. Background Feynman gauge also
helps to simplify the calculations. This method is applicable to any
Feynman diagram with any number of loops as long as the external
masses
can be ignored, and it is at least as efficient
as the string technique in the special circumstances when the latter
can be
used. In order to minimize the very considerable
algebra encountered in non-abelian gauge theories, graphical methods
are
developed for most of the calculations.

This enables the large number of terms encountered to be organized
visually in the Feynman diagram without the necessity of having to
write down
any of them algebraically.
A one-loop four-gluon amplitude in
a particular helicity configuration is computed explicity to
illustrate the
method.
\vfill
\leftline{*\ email address: Lam@physics.mcgill.ca}
\vfill\eject

\centerline{\bf 1. Introduction}
\bigskip

The number of diagrams and the number of terms in a QCD calculation
increase
dramatically with the multiplicity of the external particles as well
as the
number of loops.  Even in the absence of quarks, a pure QCD {\it
tree} process
describing the production of six gluon jets from a glue-glue
collision is given
by the sum of some 34,000 diagrams, and roughly half a billion terms.
A
one-loop pure QCD glue-glue elastic scattering amplitude has 39
diagrams and
some ten thousand terms.  The large number of diagrams is due to the
large
number of ways triple and quadruple gluon (and ghost) vertices can be
assembled
together, and the large number of terms is due to the presence of six
terms at
each vertex.  The necessity of having to sum over intermediate color
indices
makes it more complicated; if loops are present loop integrations
must be done
and the problem gets worse.  Similar complexity occurs in electroweak
computations.  These difficulties are not something that one can
ignore in
practice because a large number of jets is present at high energies,
and
because loop calculations are increasingly demanded for precision
comparisions
with the Standard Model.  To make progress one must find a way around
this
algebraic jungle.

Unnecessary algebraic complications are already present in QED
bremstrahlung
calculations as is evidenced by the fact that simple results emerge
from
complicated covariant technique calculations [1].  It was later
discovered that
the use of the {\it spinor helicity technique} enables one to obtain
the final
result directly in a much simpler way [1,2].  Over the last ten years
or so
this technique has been further developed and applied to various QCD
and
electroweak processes in the {\it tree approximation} [1--18].  It
leads to a
tremendous simplification in the calculations, reducing impossibly
large number
of terms into manageable sizes.  As a result of these techniques,
many tree
amplitudes which are too complicated to calculate by ordinary means
have been
successfully computed.  See Ref.~[16] for an excellent review of the
techniques
and the results.

The basic idea of this technique is that quark masses are negligible
at high
energies.  If they are neglected, chirality is conserved, and this
conservation
can be exploited to simplify the calculations.  For example, the
trace
$tr(\gamma p_1\cdots \gamma p_{2n})$ which according to the usual
formula is
given by the sum of $(2n)!/2^n(n)!$ terms, can be written using this
technique
as a sum of merely two terms if all $p_i^2=0$ (see eq.~(39) below).
This
simplification is equally applicable to processes involving gluons
and photons
if their wavefunctions are written in a multispinor basis.  Moreover,
gauge
freedom allows one to choose these wavefunctions to be orthogonal to
any
massless `reference momentum', thereby further simplifying the
calculations by
rendering many terms zero.  On top of that, recursion [4] and
supersymmetry
[17,12] relations may be exploited to further simplify the
calculations.

Unfortunately it is difficult to apply this beautiful technique to
the
computation of loop diagrams.  The internal loop momenta are
offshell;
chirality is not conserved and massless spinor methods are not useful
for these
momenta.  In an era when precision experiments are increasingly
called for this
is a serious handicap and it is important to find a way around this
obstacle.
This turns out to be possible and this is the subject matter of the
present
paper.

The idea is very simple, though the detailed implementation of the
idea is far
from being trivial.  If a representation of the scattering amplitude
can be
found where the internal loop momenta do not appear, then every
momentum in the
problem is a linear combination of the external massless momenta and
the spinor
helicity technique can be used.

This happens to be true for certain one-loop amplitudes which have a
string
extension.  The point is that a one-loop string scattering amplitude
can be
written as an integral over the Koba-Nielsen variables, without the
explicit
presence of any internal momenta.  By taking the string tension to
infinity,
one obtains a formula for the scattering of the massless particles in
the
string in which no internal momentum is present and the spinor
helicity
technique can be used.  The simplification thus achieved is very
considerable
[19--21].

This ingenious method has its limitations.  It is difficult to find a
simple
string formula beyond one loop, and in any case the technique is not
applicable
if the corresponding field theory has no string extension, or that
the one-loop
formula for the desired external particle is difficult to write down,
as is the
case for external fermions.  Moreover, one cannot help but feel that
there must
be a purely field-theoretical way of calculating a field-theoretical
scattering
amplitude, without having to resort to the artificial, though
ingenious, means
of creating an intermediate string and then destroying it again by
taking the
infinte tension limit.

There is indeed a well-known and purely field-theoretical way of
getting rid of
the internal loop momenta:  by introducing the Feynman parameters to
combine
the propagators, the loop-momentum integrations can be explicitly
carried out.
The result is again a formula in which the only momenta present are
the
external massless momenta, and thus the spinor helicity technique can
again be
used.  It is this route that we would like to explore here.  Unlike
the string
technique, Feynman parameter representations are available for any
Feynman
diagram with any number of loops, so this method can be used for all
processes
subject only to the validity of ignoring the external masses.  As a
matter of
fact, in the known cases [19--21], the Koba-Nielsen parameters reduce
themselves to the Feynman parameters in the infinite tension limit of
the
string, thus suggesting the close connection of the two methods.

This simple idea must be supplemented by a number of other
developments to make
it useful, for otherwise the amount of algebra necessary to carry out
the
calculations using elementary means is too unmanageable.  These means
are
available; they will be mentioned below and discussed in much more
detail in
the next two sections.

First of all, one needs a set of rules analogous to the usual
momentum-space
Feynman rules to write down the Feynman parameter representation {\it
directly
from the Feynman diagram}.  Otherwise if we had to do the internal
loop
integrations explicitly every time then the task would be too
complicated.
These Feyman-parameter rules are available [22] and will be reviewed
in Sec.~2.

The number of terms in the Feynman-parmeter representation is even
larger than
the number of terms in the conventional momentum representation.
This would
not have represented progress towards simplification except for the
fact that
the spinor helicity technique is now available to render many terms
zero.  But
to be able to do that one must first find a way to organize the large
number of
terms in a simple and systematic way so that one can recognize
beforehand which
terms to discard in the calculation.  The way to do that is to
reorganize these
terms into gauge-invariant {\it color subamplitudes}.  It is known
how this can
be done algebraically in tree and one-loop processes [16,19].  We
shall discuss
in Sec.~3 how this can be one for any number of loops {\it
graphically} by
introducing {\it color-oriented} Feynman diagrams.  In this graphical
language
the different terms in the scattering amplitude correspond to
different {\it
covering paths}.  The use of this graphical language does not in any
way reduce
the number of terms, but it gives a way to organize them in a visual
way
without the necessity of writing down anything algebraically.  This
graphical
organization can be used in the usual momentum-space representation
of a
scattering amplitude, as well as the Feynman-parameter space
representation
discussed in Sec.~2.

This graphical technique is particularly useful when it is combined
with the
spinor helicity technique, which as a result of chirality
conservation renders
many terms zero.  In graphical language this means that paths of
certain
topologies lead to vanishing results and do not have to be included.
We shall
also find that the use of Feynman gauge in the background field
method reduces
further the amount of labour of calculation by rendering more terms
zero.  A
review of the spinor helicity technique and how this can be
implemented
graphically will be discussed in Sec.~4.

We choose to illustrate the present method in Sec.~5 by computing the
one-loop
gluon-gluon elastic scattering amplitude in a particular helicity
configuaration.  This amplitude in the absence of quark loops has
already been
computed in the string method [19--21]; we choose it to illustrate
our method
so that the efficiency of the two techniques can be compared.  We
shall find
that within the present framework there are two ways to compute this
amplitude.
The direct way yields a result as efficient as the string method; the
indirect
way making use of supersymmetry is even simpler and the result can be
obtained
in only a few lines.  This is to be compared with the ordinary
Feynman diagram
calculations where some ten thousand terms appear.

It should be emphasized that chirality conservation affects only the
spin
flows, {\it viz.,} the derivative couplings and the numerators of the
propagators in the usual momentum-space representation.  The
denominators of
the propagators may remain massive without in any way affecting the
effectiveness of these techniques.  This means that while the
external
particles must remain massless, exchange and internal particles may
often be
massive, as is the case for the $Z$ and $W$ bosons.  Thus the present
technique
can be used to compute heavy particle productions if their subsequent
decays
into light particles are also incorporated into the diagrams.

\bigskip \bigskip

\centerline{\bf 2. Feynman-parameter Rules}
\bigskip
Consider a Feynman diagram in $d$-dimensional spacetime, with $N$
internal
lines and $\ell$ loops.  Let $p_A$ be the external {\it outgoing}
momenta,
$q_r\ (r=1,\cdots,N)$ be the momentum of the $r$th propagagor, and
$m_r$ be the
mass of the particle being propagated.  Every $q_r$ is given by a
linear
combination of $p_A$ and the $\ell$ loop-momenta $k_a$, the specific
combination depends on the topology of the diagram.

The scattering amplitude corresponding to a Feynman diagram expressed
in
momentum-space representation is of the form $${\cal
M}=\left({-i\over
(2\pi)^4}\right)^\ell \int\prod_{a=1}^\ell(d^dk_a){S(q,p)\over
\prod_{r=1}^N(-q_r^2+m_r^2-i\epsilon)},\eqno(1)$$ where $S(q,p)$
receives its
contributions from the vertices and the numerators of propagators.
It also
contains the symmetry factor and the minus sign for each fermion
loop.  By
introducing the Feynman parameters $\alpha_r$ and carrying out the
integrations
over $k_a$, it can be shown [22] that $${\cal
M}=\left({\pi^{d/2-4}\over
16}\right)^\ell \sum_{k=0}\Gamma\left(N-{d\ell\over 2}-k\right)\int
{{\cal D}^N
\alpha \over \Delta(\alpha )^{d/2}} { S_k(J,p) \over D(\alpha
,p)^{N-d\ell/2-k}}\equiv\sum_{k=0}{\cal M}_k,\eqno(2)$$ where
$$\eqalignno{
{\cal D}^N \alpha =&(\prod_{r=1}^N d\alpha _r) \delta (\sum_{r=1}^N
\alpha
_r-1),&(3)\cr \Delta(\alpha )=&\sum_{T_1}(\prod^\ell \alpha ),&(4)\cr
D(\alpha
,p)=&\sum_{r=1}^N \alpha _rm_r^2-P(\alpha ,p),&(5)\cr P(\alpha
,p)=&\Delta
(\alpha )^{-1}\sum_{T_2}(\prod^{\ell+1}\alpha ) (\sum^\ell
p)^2,&(6)\cr
J_r=&\Delta (\alpha )^{-1}\sum_{T_2(r)}\alpha
_r^{-1}(\prod^{\ell+1}\alpha
)(\sum^\ell p),&(7)\cr S_0(J,p)=&S(J,p).&(8)\cr}$$

The quantities appeared in (2) have a very simple physical
interpretation.  If
we consider the Feynman diagram as an electrical circuit, with the
external
momenta $p_A$ as the external currents, and the Feynman parameters
$\alpha _r$
as the resistance of the $r$th line, then $J_r$ is simply the current
flowing
through the $r$th line, and $P(\alpha ,p)$, which can be proven to be
equal to
$\sum_{r=1}^NJ_r^2 \alpha _r$, is just the power dissipated in the
circuit.
The cryptic formulas (3)--(8) offers a simple and practical way to
compute
these currents and the power directly from the Feynman diagram.

We shall now elaborate on these cryptic formulas.  A connected
diagram with
$\ell$ loops can be made into a connected tree diagram if $\ell$
internal lines
are cut.  There are many ways to do this, each resulting in a
different
(one-)tree $T_1$.  The sum in (4) is taken over all such one-trees
$T_1$, with
each term in the sum equal to the product of all the Feynman
parameters $\alpha
$ of the cut lines.  As a result, $\Delta(\alpha)$ is of a
homogeneous degree
$\ell$ in the $\alpha $'s.

Similarly, $\ell+1$ cuts can bring the diagram into two connected
trees that
are disjoint, or a `two-tree' $T_2$.  The sum in (6) is over the set
of all
such two-trees $T_2$.  This time each term consists of the product of
the
$\ell+1$ Feynman parameters of the cut lines, times the square of the
sum of
all the external momentum $p_A$ attached to one of the two trees.  It
does not
matter which tree we choose to compute the momentum sum because of
conservation
of momentum.

Now we come to the numerator $S_k(J,p)$ in (2), for $k=0,1,2,\cdots$.
The
first term $S_0(J,p)$ is just the numerator factor $S(q,p)$ in (1)
with each
$q_r$ replaced by $J_r$.  The rule for computing the current $J_r$ is
given in
(7), where the sum is over the set of all two-trees $T_2(r)$ obtained
by having
the line $r$ always cut.  The summand consists of the product of the
$\alpha$'s
of the cut lines {\it except} the $r$th (so it is of homogeneous
degree $\ell$
in $\alpha$), times a momentum factor given by the sum of all the
external
momenta attached to one of the two resulting trees.  If the momentum
$q_r$
flows from tree 1 to tree 2, then it is the sum of $p_A$ attached to
tree 2, or
minus the sum of $p_A$ attached to tree 1, that should be used in the
sum.
This convention presumes that all the external momenta $p_A$ are
outgoing.  In
other words, the sign is such that the direction of the flow of the
current
$J_r$ must match that of the external currents $p_A$.

We are now in a position to describe $S_k(J,p)$ for $k>0$.  It is
obtained from
$S_0(J,p)$ by contracting $k$ pairs of $J$'s in all possible ways,
and summing
over all such contractions.  If no contractions are possible then
$S_k=0$.  For
each pair $J_r, J_s$ in the contraction, one makes the replacement
$$J^\mu
_rJ^\nu _s\to -{1\over 2}g^{\mu \nu }H_{rs},\eqno(9)$$ and the factor
$H_{rs}$
is given by $$\eqalignno{ H_{rr}=&-\Delta(\alpha)^{-1} \partial
\Delta (\alpha
)/\partial \alpha _r,&(10)\cr H_{rs}=&\pm\Delta (\alpha
)^{-1}\sum_{T_2(rs)}(\alpha _r \alpha _s)^{-1} (\prod^{\ell+1}\alpha
),\quad(r\not= s).&(11)\cr}$$ This time the sum in (11) is over the
set of all
two-trees $T_2(rs)$ in which lines $r$ and $s$ must have been cut,
and each
term in the sum is a product of the $\alpha $'s of the cut lines
except the
$r$th and the $s$th.  The sign in front is $+1$ if both $q_r$ and
$q_s$ flow
from tree 1 to tree 2, and $-1$ otherwise.

This concludes the description of the quantities in (2).  We shall
now
illustrate these rules with one-loop diagrams.  See Ref.~[22] for an
illustration of these rules for a two-loop diagram.

A tree is obtained from a one-loop diagram by cutting any of its $N$
internal
lines.  Thus for any one-loop diagram, (4) with (3) yields
$$\Delta(\alpha)=\sum_{r=1}^N\alpha_i=1\eqno(12)$$ and (10) gives
$$H_{rr}=-1.\eqno(13)$$ Now specialize to a box diagram (Fig.~1(a))
and a
vertex diagram (Fig.~1(b)).  Using (6), (7), (11) and (12), one gets
for the
box diagram $$\eqalignno{
P(\alpha,p)&=\alpha_1\alpha_3(p_1+p_2)^2+\alpha_2\alpha_4(p_1+p_4)^2
+\alpha_1\alpha_2p_1^2+\alpha_2\alpha_3p_2^2+\alpha_3\alpha_4p_3^2
+\alpha_4\alpha_1p_4^2,&\cr
J_1&=\alpha_2p_1+\alpha_3(p_1+p_2)+\alpha_4(p_1+p_2+p_3),&\cr
J_2&=\alpha_3p_2+\alpha_4(p_2+p_3)+\alpha_1(p_2+p_3+p_4),&\cr
J_3&=\alpha_4p_3+\alpha_1(p_3+p_4)+\alpha_2(p_3+p_4+p_1),&\cr
J_4&=\alpha_1p_4+\alpha_2(p_4+p_1)+\alpha_3(p_4+p_1+p_2),&\cr
H_{rs}&=-1\quad
(r\not= s),&(14)}$$ and for the vertex diagram $$\eqalignno{
P(\alpha,p)&=\alpha_1\alpha_2(p_2+p_3)^2+\alpha_2\alpha_3p_2^2+\alpha
_1\alpha_3
p_3^2,&\cr J_1&=-\alpha_3p_3-\alpha_2(p_2+p_3),&\cr
J_2&=\alpha_3p_2+\alpha_1(p_2+p_3),&\cr
J_3&=\alpha_1p_3-\alpha_2p_2,&\cr
H_{rs}&=-1\quad (r\not= s).&(15)}$$

Before leaving this section let us say a word about renormalization.
For
theories that are no more than logarithmically divergent, primitive
ultraviolet
divergence, if any, comes from the term with the highest power of $q$
in the
numerator of (1), and this power is $2(kmax)=2(N-2\ell)$.  In the
language of
(2), this occurs only in the term $S_{kmax}(J,p)$ with the maximum
number of
contractions.  If we let $d=4+2\epsilon$, then this term is $${\cal
M}_{kmax}=\left({\pi ^{2+\epsilon }\over 16}\right)^\ell\Gamma(\ell
\epsilon )
\int {{\cal D}^N \alpha \over \Delta (\alpha
)^{2+\epsilon}}{S_{kmax}(J,p)
\over D(\alpha ,p)^{\ell \epsilon }}.\eqno(16)$$ Renormalization in
the MS or
$\overline {MS}$ scheme is therefore easy to carry out.

\bigskip \bigskip

\def\g{{\cal G}}
\def\.{\cdot}
\def\bk#1{\langle #1\rangle}

\centerline{\bf 3. Color Decomposition and Spin Flow}
\bigskip

We specialize now to QED and $SU(N)$ QCD.  With simple modifications
this can
also be applied to the electroweak processes.  The quark (q) belongs
to the
fundamental representation and both the gluon (g) and the
Fedeev-Popov ghost
(G) belong to the adjoint representation.  For the purpose of
discussing color
decomposition, there is no need to distinguish `g' and `G' so we
shall
collectively refer to them as `$\g$'.

The factor $S(q,p)$ in (1) is composed of vertex contributions and
the
numerators of propagators.  As such it contains information on both
spin and
color.  It is this factor that contains the large number of terms
mentioned in
the Introduction.  The purpose of this section is to discuss how this
quantity
(and the corresponding scattering amplitude ${\cal M}$) can be
reorganized to
simplify calculations.  Specifically, the factor $S(q,p)$ (the
amplitude ${\it
M}$) is to be decomposed into the form $\sum_i {\cal C}_im_i$, where
${\cal
C}_i$ is the {\it color factor}, which consists of products of the
$(S)U(N)$
generators and their traces, and $m_i$ carries spin and momentum but
no color
information.  We shall refer to $m_i$ in the decomposition of
$S(q,p)$ as the
{\it spin factor} (other than some trivial factors of the coupling
constant $g$
to be discussed later), and $m_i$ in the decomposition of ${\cal M}$
as the
{\it color subamplitude}.  There are at least three advantages for
such a
decomposition.  First of all, many ${\cal C}_i$'s differ from one
another only
by permutations of the color indices.  Consequently only one of these
$m_i$'s
has to be computed explicitly; the rest of them can be obtained by
similar
permutations.  Secondly, the color factors ${\cal C}_i$ are
independent. As a
result each spin factor and color subamplitude are invariant under an
arbitrary
gauge change of an external polarization vector.  This aspect of it
will be
particularly useful in the spinor helicity technique because we can
choose the
reference momenta independently for each $m_i$.  Thirdly, the $m_i$'s
satisfy
other important identities like the `dual Ward's identity' which can
be
utilised in practical computations.  Color decomposition has been
carried out
algebraically for tree and one-loop diagrams [16,19]; we shall do it
to all
loops and do it {\it graphically} in order to minimize the algebra.
For that
purpose we will introduce {\it color-oriented Feynman diagrams}, from
which
color factors as well as spin factors can be read off directly.

The discussion in this section is independent  of the last section.
Hence
the results are equally applicable to momentum-space representations
as well as
Feynman-parameter representations.

It is convenient to extend the gauge theory by an extra $U(1)$ factor
to
complete it to an $U(N)=U(1)\times SU(N)$ gauge theory.  This
simplifies the
algebraic manipulation without losing any information, for as we
shall see QCD
($SU(N)$) expressions can be read off from the simpler results of an
$U(N)$
gauge theory.

Let $T^A\ (A=0,a; a=1,\cdots, N^2-1)$ be the $U(N)=U(1)\times SU(N)$
generators
in the fundamental representation.  $T^0=(1/\sqrt{N})${\bf 1} is the
$U(1)$
generator and $T^a$ are the $SU(N)$ generators.  The normalization of
the
$U(1)$ factor is chosen to satisfy the normalization
$$Tr(T^AT^B)=\delta
^{AB}.\eqno(17)$$ The structure constant $f^{ABC}$ can be obtained
from the
commutation relation $[T^A,T^B]=if^{ABC}T^C$ by the formula
$$f^{ABC}={-i}\{Tr(T^AT^BT^C)-Tr(T^AT^CT^B)\},\eqno(18)$$ and is seen
to be
antisymmetric in its $U(N)$ indices.  Note from this that
$f^{0BC}=0$, which
reflects the physics that the $SU(N)$ and the $U(1)$ gauge bosons do
not
interact directly with each other.  This is an important fact which
will allow
us to project out the $U(1)$ bosons to regain QCD.

The completeness relation dual to (17) is
$$(T^A)_{ij}(T^A)_{kl}=\delta
_{jk}\delta _{il},\eqno(19)$$ where summation over the $N^2$ repeated
indices
$A$ is understood.  From this one obtains
$$f^{ABE}f^{ECD}=(-i)^2\{Tr(ABCD)-Tr(BACD)-Tr(ABDC)+Tr(BADC)\}.\eqno(
20)$$ For
brevity, we have chosen to write $Tr(T^AT^BT^CT^D)$ simply as
$Tr(ABCD)$, and
we shall often use this same abbreviation of replacing $T^A$ simply
by $A$ in
the rest of this paper.

The vertices for QCD and $U(N)$ gauge theory in the background
Feynman gauge
[23] are exhibited in Fig.~2.  A thin solid line stands for a gluon,
a dashed
line stands for the Fadeev-Popov ghost, and a thick solid line stands
for a
fermion.  The background gauge is a gauge in which an external gluon
is
distinguished from an internal gluon; in Fig.~2 a circled `A' at the
end of the
line signifies an external line.  In diagrams where a circled `A'
makes an
appearance, the uncircled gluon lines are meant to be internal lines.
In
diagrams where no circled `A' appear, each of the gluon lines in the
diagram
may be taken either as an internal line or an external line, unless
such a
combination of external and internal gluon lines have appeared
explicitly in a
diagram in Fig.~2 in which a circled `A' is present. In that case the

Feynman rule (see eq.~(21) below) for the diagram with circled `A's

should be used.

The discussion in this section can be applied to any gauge.  However
for the
sake of application in the next two sections we shall use explicitly
the
background Feynman gauge.  Although there are more vertices (those
with circled
`A') in the background gauge than the usual covariant gauges,
nevertheless we
shall see in the next section that the use of background gauge along
with the
spinor helicity basis simplifies enormously the calculations.

All the momenta in Fig.~2 are understood to be pointing outwards.  It
is also
understood that the line labelled by 1 carries an outgoing momentum
$p_1$, a
color index $a$, and a Lorentz index $\alpha $.  Similarly, line 2
carries the
quantum numbers $(p_2,b,\beta )$, etc.  The Feynman rules for these
vertices in
the background {\it Feynman gauge} are given below, with the equation
numbers
corresponding to the diagrams in Fig.~2.  For example, eq.~(21a) is
the Feynman
rule for the vertex in Fig.~2(a).  \def\b{\bullet\ } $$\eqalignno{
\b&igf^{abc}\{g_{\alpha \beta }(p_1-p_2)_ \gamma +g_{\beta \gamma
}(p_2-p_3)_
\alpha+g_{\gamma \alpha }(p_3-p_1)_ \beta \}, &(21a)\cr
\b&igf^{abc}\{g_{\alpha
\beta }(2p_1)_ \gamma +g_{\beta \gamma }(p_2-p_3)_ \alpha+g_{\gamma
\alpha
}(-2p_1)_ \beta \}, &(21b)\cr \b&-g^2f^{abe}f^{ecd}(g_{\alpha \gamma
}g_{\beta
\delta }- g_{\alpha \delta }g_{\beta \gamma })-g^2f^{ace}f^{ebd}
(g_{\alpha
\beta }g_{\gamma \delta }-g_{\alpha \delta }g_{\beta \gamma }) &\cr &
-g^2f^{bce}f^{ead}(g_{\beta \alpha } g_{\gamma \delta }-g_{\beta
\delta
}g_{\alpha \gamma }),&(21c)\cr \b&-g^2f^{abe}f^{ecd}(g_{\alpha \gamma
}g_{\beta
\delta }- g_{\alpha \delta }g_{\beta \gamma }+ g_{\alpha \beta
}g_{\gamma
\delta})-g^2f^{ace}f^{ebd} (g_{\alpha \beta }g_{\gamma \delta
}-g_{\alpha
\delta }g_{\beta \gamma }) &\cr & -g^2f^{bce}f^{ead}(g_{\beta \alpha
}
g_{\gamma \delta }-g_{\beta \delta }g_{\alpha \gamma }-g_{\beta
\gamma
}g_{\alpha \delta }),&(21d)\cr \b&-igf^{abc}(-p_3)_ \alpha ,&(21e)\cr
\b&-igf^{abc}(p_2-p_3)_ \alpha ,&(21f)\cr
\b&-g^2f^{abe}f^{ecd}g_{\alpha \beta
},&(21g)\cr \b&-g^2(f^{abe}f^{ecd}+f^{ace}f^{ebd})g_{\alpha \beta
},&(21h)\cr
\b&gT^a \gamma_\alpha .&(21i)\cr}$$ The $U(N)$ Feynman rules are the
same
except that the $SU(N)$ color indices $a,b,c,d$ should be replaced by
the
$U(N)$ indices $A,B,C,D$.

We shall use (18) and (20) to replace the factors $f^{ABC}$ and
$f^{ABE}f^{ECD}$ in (21), and then proceed to group the terms with
the same
$U(N)$ traces.  Each of these terms defines a {\it color-oriented
vertex} in
which the $U(N)$ indices of the external lines read clockwise
coincides with
the indices in the trace read from left to right.

Each color-oriented vertex factor is a product of three quantities:
the
coupling constant factor, the color factor, and the spin factor.  The
coupling
constant factor will be taken to be $g$ for cubic vertices and $g^2$
for
quartic vertices.  The color factors will be taken to be $T^A$ for a
qqg
vertex, to be $Tr(ABC)$ for a $\g\g\g$ vertex, and to be $Tr(ABCD)$
for a
$\g\g\g\g$ vertex.  The rest of the vertex factor will be defined to
be the
{\it spin factor}, the details of which are exhibited in Fig.~3.

Fig.~3 should be read in the following way.  A gluon line continuing
through
the vertex indicates a factor of a metric tensor in spacetime.  A dot
represents the vector written below the diagram; other numerical
factors are
also written below the diagram.  For example, the color factor for
diagram (a)
is $Tr(ABC)$, its coupling-constant factor is $g$, and its spin
factor is
$g_{\beta\gamma}(p_2-p_3)_\alpha$.  The color factor for diagram (e)
is
$Tr(ABCD)$, its coupling-constant factor is $g^2$, and its spin
factor is
$2g_{\alpha\gamma}g_{\beta\delta}$.

The Feynman diagrams assembled from color-oriented vertices will be
called {\it
color-oriented Feynman diagrams}, or just an {\it oriented diagrams}
for short.
A color-oriented diagram can be obtained from an ordinary Feynman
diagram by
flipping any number of external gluon lines each about the $\g$
propagator it
emerges from, by interchanging two identical external $\g$ lines
emerging from
the same vertex, by interchanging two identical internal $\g$ lines
if this
does not alter the topology of the diagram, or a combination of
these.  In
general, an ordinary Feynman diagram leads to many color-oriented
Feynman
diagrams.  The total contribution to a scattering amplitude is the
sum of the
contributions from all the color-oriented diagrams.

If the Feynman diagram in question can be obtained from the infinite
tension
limit of a string diagram, then flipping of the gluon line
corresponds to a
twisting of the string.  The color factors for the whole diagram to
be
discussed below are nothing but the Chan-Paton factors [24].

The total color factor of an oriented diagram is the product of the
color
factors of its oriented vertices, summed over intermediate color
indices.
Eq.~(19) can be used to carry out these sums; the result of which
gratifyingly
can be read off once again directly from the color-oriented Feynman
diagram.

Let us start with a tree diagram having $n$ external $\g$ particles
and no
fermion anywhere in the diagram.  The color factor of this tree turns
out to be
$$Tr(C_1C_2\cdots C_n),\eqno(22)$$ where $C_1,C_2,\cdots,C_n$ are the
$U(N)$
color indices of the oriented Feynman diagram read {\it clockwise
around} the
whole tree.  For example, the color factor for Fig.~4 is \break
$Tr[(1)(2)(3)(4)\cdots(14)(15)(16)]$.

{}From now on, we shall use capital letters near the end of the
alphabets to
denote products of $U(N)$ generators, {\it e.g.}, $X=C_1C_2\cdots
C_p$.

Eq.~(22) can be proven by induction.  By definition, a color-oriented
diagram
with a single $\g\g\g$ or a single $\g\g\g\g$ vertex is already of
the form of
(22).  Suppose now we have two trees, the color factor of each is of
the form
(22).  This is illustrated in Fig.~5, where the color factors of the
two trees
are respectively $Tr(XA)$ and $Tr(AY)$.  When we sew these two trees
together
at index `A' to obtain a bigger tree, the resulting color factor,
using (19),
is indeed $Tr(XY)$, which can be read out directly from Fig.~5 using
the rules
of (22).  This completes the induction proof of (22).

The result for $SU(N)$ QCD is equally simple and the color factor is
again
given by (22), but with the upper case $U(N)$ indices $C_i$ replaced
by the
corresponding lower case $SU(N)$ indices $c_i$.  This is so because
of the
absence of coupling between the $U(1)$ and the $SU(N)$ gauge bosons,
{\it
viz.,} $f^{0BC}=0$.  Therefore as long as the external lines of a
connected
tree carry an $SU(N)$ indices, the $U(1)$ gluon is decoupled and will
never
makes its appearance in the internal line either.

Next, consider tree diagrams in which a single fermion line is
present.  The
color factor for a qqg vertex is $T^A$ if $A$ is the $U(N)$ color
index of the
gluon.  If a whole tree of $\g$ particles with the color factor
$Tr(XA)$ is
planted at this vertex, then the combined color factor is obtained
from (19) to
be $X$.  If $A_1,A_2,\cdots,A_n$ are the successive qqg vertices as
we go along
a fermion line, and if $\g$-trees with color factors $Tr(X_iA_i)$ are
planted
at these vertices, then the combined color factor would be
$$X_1X_2\cdots
X_n.\eqno(23)$$ Graphically, this is simply the the multiplication of
all the
$U(N)$ generators $T$ in clockwise order around the whole tree, as
shown in
Fig.~6.

Note the difference between a ggg and a qqg vertex.  The former is
oriented, in
the sense that there are two oriented vertices associated with one
ordinary
Feynman vertex, but the oriented vertex in the latter case is the
same as the
ordinary vertex.  In the ggg vertex, there is no further color
specification
other than the indices of the external $\g$-lines, but in the qqg
vertex, the
color state of the initial and the final quarks must still be
specified.  The
result of these differences is that as we traverse clockwise around
an oriented
diagram to read out its color factor, we must cover {\it both sides}
of every
$\g$-line, and that the {\it trace} of the product of generators must
be taken.
On the other hand, we should traverse only through the {\it top side}
of a
fermion line and {\it no trace} of the product of generators is to be
taken.
If we should find it convenient to draw a $\g$-tree downward from a
fermion
line, as is the case of the $X_2$-tree in Fig.~6, then clockwise
order must
still be maintained in the way indicated in the figure.  In other
words, since
we only follow the top and not the bottom of the fermion line, the
color factor
in Fig.~6 is $X_1X_2 X_3$, and not $X_1X_3X_2$ as we might think if
we were to
follow both sides of the fermion line.

The same result (23) is again true if we consider only $SU(N)$ QCD.
Once again
this is due to the lack of coupling between the $U(1)$ and the
$SU(N)$ gluons.

The situation of having a $\g$-tree connecting two separate quark
lines can be
obtained similarly from (19), but the paths along which the color
generators
are multiplied together now cross over from one fermion line to
another, as in
Fig.~7.  This is so because $$Tr(AYBV)XAW\otimes UBZ=XYZ\otimes
UVW.\eqno(24)$$
If on the way one encounters another $\g$-tree connecting to a third
fermion
line, then one must cross over to the third tree at that point, etc.

This formula is still valid in $SU(N)$ QCD as long as either $Y$ or
$V$ factor
appearing in the tree $Tr(AYBV)$ connecting the two fermion lines
contains at
least one external $SU(N)$ gluon.  In that case, as before,
decoupling prevents
the $U(1)$ gluon to appear even in the internal lines.  The situation
is
different if the tree is simply $Tr(AB)$.  In that case an $U(1)$
gluon
connecting the two fermion lines is present, and its effect must be
subtracted
away when $SU(N)$ is considered.  The result is then $$XY\otimes
UW-(1/N)XW\otimes ' UZ.\eqno(25)$$ The second term follows the
original fermion
lines all the way without a cross-over, and this distinction from the
first
term is indicated in the formula by using $\otimes'$ rather than
$\otimes$.

Loop diagrams are obtained by joining ends of tree diagrams.
Consider first
the cases when ends of fermions are joined into fermion loops.  In
the presence
of a single fermion line, or whenever such a fermion line is not
attached to
another fermion line by gluons, then all we have to do is to take the
trace
over (23).  If two fermion lines are present, as in Fig.~7, and if
the two ends
of the top fermion line are joined together to form a fermion loop,
then the
color factor for an $U(N)$ gauge theory can be obtained from (24) to
be
$$XYZUVW.\eqno(26)$$
Note that this can be read off directly from Fig.~7 as long as we
remember to cross over at the $\g$-tree.  Other cases involving more
fermions
can be obtained similarly.

Consider next a fermion-$\g$ loop, as in Fig.~8, obtained by
attaching a
$\g$-tree of color factor $Tr(AYBU)$ to a fermion line with color
factor
$XAVBZ$ at points $A$ and $B$.  Again first imagine point $A$ to have
been
attached but not point $B$.  Then the color factor of the combined
tree is
$XYBUVBZ$, summed over $B$.  This yields $$XYZTr(UV).\eqno(27)$$ Note
that this
factor can again be read off directly from Fig.~8:  $XVZ$ is the
multiplication
of the color generators clockwise order following the outside path of
the loop,
and $Tr(VY)$ is the trace factor corresponding to the lines inside of
the loop.
This feature about tracing the outside of a loop and the inside
separately will
occur again when we discuss $\g$-loops.  Note that the outside of the
loop,
like the original trees, is followed clockwise, whereas the inside of
the loop
is followed counter-clockwise.

As a check, note that Fig.~8 can also be obtained from Fig.~7 by
joining the
right end of the bottom fermion line to the left end of the top
fermion line.
In this way one again obtains (26).

Lastly, we will consider sewing ends of a $\g$-tree together to form
a loop, as
in Fig.~9.  Before we fuse it together at the point $A$, the color
factor for
the tree is $Tr(RSAXYATU)$.  Summing over $A$ yields
$$Tr(RSTU)Tr(XY).\eqno(28)$$ The result can be read again directly
from the
graph.  The first trace is taken in clockwise order along a closed
path around
the whole diagram passing through the outside of the loop; and the
second trace
is taken in counter-clockwise order along the closed path passing
through the
inside of the loop.

One can apply (19) to more complicated diagrams with any number of
loops.  The
result in the case of an $U(N)$ gauge theory can always be read off
simply from
the color-oriented diagram.  The general rule for the color factor
${\cal C}_i$
is the following.  Circle around the diagram with continuous `{\it
color
paths}' of the following kind.  These paths may start from one end of
a fermion
line and end at another end (of possibly another fermion line), or
else they
must be closed.  The {\it upper} side of every fermion line and {\it
both}
sides of every gluon and ghost line must be covered once and only
once by these
paths.  Associate each external gluon with $U(N)$ color index $A$ the
generator
$T^A$.  Go along the path in clockwise order (counter-clockwise order
if it is
inside a loop) and multiply these generators successively from left
to right.
If the path is an open path, this product is the color factor
associated with
the path.  If the path is closed, then a trace should be taken.  The
total
color factor for the color-oriented connected diagram is the product
of these
individual color factors.

See Figs.~4--9 for illustrations.

The rules for $SU(N)$ can be obtained from the $U(N)$ rules by
subtracting out
the $U(1)$ gluons which remains coupled in the diagram.

Having thus a graphical way to read out the total color factor for an
oriented
diagram, the next task is to find an equally simple and general
graphical
method to read out the total spin factor of the oriented diagram.
This can be
done very simply, and the notation adopted in Fig.~3 is actually
designed with
this in mind.

To do so, cover the maximal gluon subdiagram of the oriented diagram
in
question with `{\it spin-flow paths}'.  A spin-flow path is different
from a
color path discussed above in that it stays right on the gluon lines
of the
diagram and not above or below them.  A spin-flow path is meaningful
only for a
gluon line, internal or external, and it is simply a continous path
tracing
through a portion of the gluon subdiagram.  Such a path may be a
closed path,
or an open path.  If it is an open path, it must end at an external
gluon line,
or a cubic vertex.  Conversely, there must be one and only one path
ending at
each cubic vertex.  See Figs.  12 and 13 for examples of these paths.

The spin factor associated with a closed path is $g^\mu_\mu=d$, and
the spin
factor associated with an open path is the dot product of the vectors
at the
two ends.  The vector at a cubic vertex is given in Fig.~3, and the
vector
associated with an external gluon line is simply its polarization
vector
$\epsilon$.  The total spin factor of the oriented diagram is the
product of
the spin factors of all the paths, times whatever extra numerical
factors
appearing at the quartic vertices in Fig.~3, times products of the
numerators
of fermion propagators ($\gamma\.q$) if present, summed over all
possible
spin-flow path coverings of the maximal gluon subgraph.

The coupling constant factor for an oriented diagram is simply the
product of
the coupling constant factors of all its vertices.

The numerator factor $S(q,p)$ in (1) for a Feynman diagram is then
the product
of the color factor, the coupling-constant factor, and the spin
factor, summed
over all spin-flow paths and all color-oriented diagrams.  Extra
factors such
as the minus sign associated with each closed fermion loop and the
symmetry
factor will be absorbed into $S(q,p)$ as well.  To be sure, there are
many
terms present for a complex diagram corresponding to many spin-paths
and many
color-oriented diagrams.  In fact, all that we have done up to this
point is to
give a graphical interpretation of every term that appears in
$S(q,p)$.  We
have not reduced the number of terms there in any way.  However, this
graphical
approach helps to organize the terms mentally without having to write
down a
single algebraic formula, so it helps to keep us away from the
algebraic
jungle.  The real simplification comes in only when we start using
the spinor
helicity technique and the background gauge in the next section.

Simplifications can also result from supersymmetry.  Consider for
example an
oriented diagram containing a quark line.  The spin factor is
essentially the
same when the quark is replaced by a gluino.  Under such a
replacement, the
color factor changes simply by having traces taken over the original
color
factor.  So the colored subamplitudes of a quark diagram is the same
as that
for a gluino diagram.  On the other hand, the colored subamplitude of
a gluino
diagram is related to that for a pure gluon diagram by supersymmetry.
This
chain of reasoning makes it possible to related pure gluon amplitudes
with
those with a quark line in it.  Such supersymmetry relations [16,17]
have been
used in tree processes to simplify calculations, and as we shall see
in Sec.~5,
it can be used to simplify calculations for loop amplitudes as well.

\bigskip\bigskip
\centerline{\bf 4. Spinor helicity basis}
\bigskip

We have discussed how to organize graphically the numerator factor
$S(q,p)$ in
the last section.  Nevertheless, there are many terms involved,
corresponding
to the many color-oriented diagrams and the many spin-flow paths for
a given
diagram.  A clever choice of gauge and polarization vectors at this
point can
render many of the terms zero, making it unnecessary to consider some
path
coverings and/or color-oriented diagrams, and thereby reduce the
labour of
computation enormously.  We shall see that the use of background
gauge together
with polarization vectors chosen in the helicity spin basis will
accomplish
this purpose.

We shall first summarize the known results of the spinor helicity
basis taken
from Ref.~[16], and then go on to discuss further simplifications
brought about
by the use of the background gauge.

Let $|p\pm\rangle$ be the incoming wave function of a massless
fermion with
momentum $p$ and chirality $\pm 1$, normalized in such a way that
$$\langle
p\pm|\gamma_\mu|p\pm\rangle=2p_\mu.\eqno(29)$$ From chirality
conservation, one
gets $$\bk{p\pm|q\pm}=0\eqno(30)$$ valid for any other massless
momentum $q$.
This is the central relation that leads to much of the
simplifications.  Let
$$\eqalignno{ \bk{pq}&=\bk{p-|q+}=-\bk{q-|p+}=-\bk{qp},&\cr
[pq]&=\bk{p+|q-}=-\bk{q+|p-}=-[qp].&(31)}$$ Then $$\eqalignno{
\bk{pq}^*&=sign(p\.q)[qp],&(32)\cr \bk{pq}[qp]&=2(p\.q),&(33)\cr
\bk{p\pm|\gamma_{\mu_1}\cdots\gamma_{\mu_{2n+1}}|q\pm}&=
\bk{q\mp|\gamma_{\mu_{2n+1}}\cdots\gamma_{\mu_1}|p\mp},&(34)\cr
\bk{p\pm|\gamma_{\mu_1}\cdots\gamma_{\mu_{2n}}|q\mp}&=
-\bk{q\pm|\gamma_{\mu_{2n}}\cdots\gamma_{\mu_1}|p\mp},&(35)\cr
\bk{AD}\bk{CD}&=\bk{AD}\bk{CB}+\bk{AC}\bk{BD},&(36)\cr
\bk{A+|\gamma_\mu|B+}\bk{C-|\gamma^\mu|D-}&=2[AD]\bk{CB},&(37)\cr
\gamma
p&=|p+\rangle\langle p+|+|p-\rangle\langle p-|.&(38) }$$

To illustrate how massless momenta and the ensuing chirality
conservation can
simplify calculations, consider the calculation of $Tr[(\gamma
p_1)(\gamma
p_2)\cdots(\gamma p_{2n-1})(\gamma p_{2n-1})]$, where every $p_i$ is
massless.
Using usual formulas, this is given by a sum of $(2n)!/2^nn!$ terms,
each
containing a product of $n$ pairs of momentum dot products.  Using
(38), (30)
and (31), this can be reduced to just a sum of two terms:
$$\bk{p_1p_2}[p_2p_3]\cdots[p_{2n-1}p_{2n}]\bk{p_{2n}p_1}+
[p_1p_2]\bk{p_2p_3}\cdots\bk{p_{2n-1}p_{2n}}[p_{2n}p_1].  \eqno(39)$$

The polarization vector for an outgoing photon or gluon with momentum
$p$ and
helicity $\pm 1$ can chosen in a multispinor basis to be
$$\epsilon^\pm_\mu(p,k)=\pm{\langle p\pm|\gamma_\mu|k\pm\rangle \over
\sqrt{2}\langle k\mp|p\pm\rangle},\eqno(40)$$ where the {\it
reference
momentum} $k$ in (40) is massless but otherwise arbitrary.  The
choice of
different $k$ corresponds to the choice of a different gauge, and
these
different choices are related by
$$\epsilon^+_\mu(p,k)\to\epsilon^+_\mu(p,k')-\sqrt{2}
{\bk{kk'}\over\bk{kp}\bk{k'p}}p_\mu.\eqno(41)$$

These polarization vectors satisfy the following identities:
$$\eqalignno{
\epsilon^\pm_\mu(p,k)&=(\epsilon^\mp_\mu(p,k))^*,&(42)\cr
\epsilon^\pm(p,k)\.p&=\epsilon^\pm(p,k)\.k=0,&(43)\cr
\epsilon^\pm(p,k)\.\epsilon^\pm(p,k')&=0,&(44)\cr
\epsilon^\pm(p,k)\.\epsilon^\mp(p,k')&=-1,&(45)\cr
\epsilon^\pm(p,k)\.\epsilon^\pm(p',k)&=0,&(46)\cr
\epsilon^\pm(p,k)\.\epsilon^\mp(k,k')&=0,&(47)\cr
\epsilon^+_\mu(p,k)\epsilon^-_\nu(p,k)+\epsilon^-_\mu(p,k)\epsilon^+_
\nu(p,k)
&=-g_{\mu\nu}+{p_\mu k_\nu+p_\nu k_\mu\over p\.k},&(48)\cr \gamma
\.\epsilon
^\pm(p,k)&=\pm{\sqrt{2}\over \bk{k\mp|p\pm}} (|p\mp\rangle\langle
k\mp|+|k\pm\rangle\langle p\pm|).&(49)\cr }$$

This completes the summary of the properties of the spin-helicity
basis.  The
vanishing dot products (43), (44), (46), (47) are what make this
basis
particularly useful.

Background gauge is convenient for loop calculations because (43) --
(47) can
be used to eliminate many terms in this gauge.  In this gauge, two of
the three
terms in the ggg vertex with an external line (see Fig.~3) involves
only the
momentum of the external line.  This enables many terms to vanish as
we shall
see in the following illustration.

Consider an $n$-gluon color-oriented diagram where either all the
gluons have
the same helicity, or all but one have the same helicity.  Let us
consider the
latter, and assume gluon 1 to have a negative helicity while all the
other
gluons have positive helicities.  Let us choose the reference vectors
$k_i$ for
the polarization vector $\epsilon (p_i,k_i)$ to be $k_1=p_2$ and
$k_i=p_1$ for
all $i\not= 1$.  This choice is designed so that (43) to (47) can be
used to
show that $$\eqalignno{ \epsilon _i\.\epsilon _j&=0,\ \ (\forall
i,j),&(50)\cr
p_i\.\epsilon_i&=p_1\.\epsilon _i=0\ \ (\forall i),&(51)\cr
p_2\.\epsilon
_1&=0.&(52)}$$

Eq.~(50) is particularly useful.  The spin factor consists of
products of dot
products of the form $\epsilon \.\epsilon ',\ \epsilon \.q,\ q\.q'$,
of degree
$n$ in the polarization vectors and of degree $m$ in the momenta if
there are
$m$ $\g\g\g$ vertices.  For tree diagrams $m\le n-2$, so at least one
$\epsilon\.\epsilon'$ must be present in every term.  Because of (50)
the tree
amplitude with this helicity configuration must vanish [16].  For one
loop
diagrams, $m=n$ only if no quartic vertices are present.  Otherwise
$m<n$, and
the corresponding contribution again vanishes on account of
$\epsilon\.\epsilon'$.  This greatly simplifies calculations because
no gggg
nor ggGG vertices need ever be considered.  Moreover, in the
Feynman-parameter
representation, contraction (9) again leads to the presence of
$\epsilon\.\epsilon'$ so current contractions never have to be
considered.  As
a result, all $S_k(J,p)=0$ except $S_0(J,p)=S(J,p)$.

Other simplifications can be seen in Fig.~10.  Paths A and B vanish
{\it in the
background Feynman gauge} because of (51) and Figs.~3(c,d).  Path C
vanishes
because of (50).  As a corollary paths between two dots like $C'$ are
also
forbidden because a path C must then be present to take up the
leftover
$\epsilon$ factors.  Path D vanishes because of (52).

\bigskip \bigskip
\centerline{\bf 5.
 One-loop four-gluon amplitude}
\bigskip

\def\f{{1\over 2}}

\def\g{{\cal G}}
\def\.{\cdot}

To illustrate techniques developed in the last three sections, we
compute in
this section a one-loop four-gluon amplitude in which three of the
four gluons
have the same helicity.  We will first compute the case when quarks
are absent
because that pure gluon amplitude has been computed with the string
technique
[19-21], so a comparison of the efficiency of the two methods can be
made.  We
find that the present method is every bit as efficient as the string
technique.

Next, we shall compute the same four-gluon amplitude with an internal
{\it
quark} loop.  Thanks to chirality conservation the calculation is
even simpler
than the pure gluon case.  Using supersymmetry arguments, this
amplitude can be
related to the pure gluonic amplitude, thereby providing a second
method to
compute the pure QCD four-gluon amplitude.  The result agrees with
the first
calculation, but the number of steps needed to reach the result is
now even
smaller.

The reference momenta for the polarization vectors will be chosen as
in the
last section, so that eqs.~(50) -- (52) can be used.  As discussed
there,
quartic vertices do not contribute, and current contractions cannot
occur so
$S_k(J,p)=0$ in (2) for $k>0$.

There is a further simplification when four-point amplitudes are
considered.
Each polarization vector is perpendicular to two momenta:  its own
gluon
momentum and its reference momentum.  That means that its dot product
with the
other two external momenta are equal and opposite, thereby resulting
in only
one independent dot product per polarization vector.  Let
\def\r{\sqrt{2}}
$$\eqalignno{ A_1&=\epsilon _1\.p_3=-\epsilon
_1\.p_4=-{\bk{13}[32]\over\r[21]},&\cr A_2&=\epsilon
_2\.p_3=-\epsilon
_2\.p_4=-{[24]\bk{41}\over\r\bk{12}},&\cr A_3&=\epsilon
_3\.p_2=-\epsilon
_3\.p_4=-{[34]\bk{41}\over\r\bk{13}},&\cr A_4&=\epsilon
_4\.p_2=-\epsilon
_4\.p_3=+{[42]\bk{21}\over\r\bk{14}}.&(53)\cr}$$ Then the numerator
$S_0(J,p)=S(J,p)$ in (2) is proportional to $A_1A_2A_3A_4$.

With quartic vertices out of the way, the color-oriented diagrams
contributing
to this process are the box diagrams, the vertex insertion diagrams,
and the
self-energy insertion diagrams.  We shall see that the self-energy
diagrams are
zero, and three out of the vertex-insertion diagrams also do not
contribute.

To see that, consider the spin-flow path of Fig.~11(a) originating
from gluon
4.  Because of eq.~(50) this path cannot end at another external
gluon line.
It cannot end at the vertex joining lines 2 and 3 either for then one
is forced
to have the factor $\epsilon _2\.\epsilon _3$, which is zero.  The
only other
possibility then is for the spin-flow to end at a vertex within the
loop, in
which case a factor $\epsilon _4\.J$ will result, where $J$ is some
combination
of the currents flowing through the loop.  From (7), we see that $J$
is a
linear combination of $p_1$ and $p_4$.  Since $\epsilon
_4\.p_1=\epsilon
_4\.p_4=0$, these paths are not allowed either.  Consequently diagram
11(a)
makes no contribution.  The same argument will hold if instead of 4
and 1 it is
gluons 1 and 2 which are attached to the loop.  Consider now diagram
11(b),
which we claim also makes no contribution.  To see that, consider a
spin-flow
path that starts from gluon 1.  This path cannot flow on to gluon 2,
so it
either ends at its own vertex, or goes beyond.  In the former case
the
contribution vanishes because $\epsilon _1\.p_1=\epsilon _1\.p_2=0$.
In the
latter case, the path starting from gluon 2 must end at its own
vertex, and
this vanishes because $\epsilon _2\.p_1=\epsilon _2\.p_2=0$.
Consequently,
three of the four vertex-insertion diagrams associated with the color
factor
$Tr(abcd)$ gives no contributions.

Essentially the same argument also shows that none of the self-energy
insertion
graphs like 11(c) makes any contribution.  This leaves only the box
graph
Fig.~12 and the vertex insertion graph Fig.~13.

Let us consider the allowed spin-flow paths in the box diagrams,
Fig.~12,
remembering that paths A,B,C,C$'$,D of Fig.~10 may not be present.
This means
the path starting from gluon 1 must end at the same vertex, or else
there must
be another path ending at vertex 1 giving rise to some
$\epsilon_i\.p_1=0$.
There are actually altogether nine possible spin flows in an internal
gluon
loop and two more in a ghost loop, as shown in Fig.~12.  Using (2)
and (14),
the amplitude from the box diagram contributing to the color factor
term
$Tr(abcd)Tr({\bf 1})$ is $${\cal M}^B={1\over 16\pi^2}\int D^4 \alpha
{S^B(J,p)\over (\alpha _2 \alpha _4s+\alpha _1 \alpha
_3t)^2},\eqno(54)$$ with
$t=(p_1+p_2)^2$, $s=(p_1+p_4)^2$, and $u=(p_1+p_3)^2$.  The numerator
is given
by $S^B(J,p)=Tr(abcd)Tr({\bf 1})g^4B$, with $B$ being the spin factor
from all
the box diagrams.  Using the spin factors for the color-oriented
vertices in
Fig.~2, and the expression for the currents in eq.~(14), one gets
$$\eqalignno{
B_1&=g^\mu_\mu[\epsilon_1\.(J_2+J_1)][\epsilon_2\.
(J_3+J_2)][\epsilon_3\.(J_4+J_3)][\epsilon_4\.(J_1+J_4)]/A_1A_2A_3A_4
&\cr
&=4(2\alpha_4)(2\alpha_4)(-2\alpha_1-2\alpha_2)(2\alpha_3)
=-64\alpha_4^2\alpha_3(\alpha_1+\alpha_2),&\cr
B_2&=[\epsilon_1\.(J_2+J_1)][\epsilon_2\.(-2p_4)][\epsilon_3\.(-2p_2)
][
\epsilon_4\.(-2p_3)]/A_1A_2A_3A_4&\cr
&=(2\alpha_4)(2)(-2)(2)=-16\alpha_4.&\cr}$$ Similar calculations show
that
$$\eqalignno{ B_3&= B_2,&\cr B_4&=B_5=16\alpha_4^2,&\cr B_6&=B_7=
16\alpha_4(\alpha _1+\alpha _2),&\cr B_8&=B_9= 16\alpha_4 \alpha
_3,&\cr
B_{10}&=B_{11}=-B_1/4.&(55)\cr}$$

The sum is $$B=\sum_{i=1}^{11}B_i=-32(\alpha _1+\alpha _2)\alpha
_3\alpha _4^2.
\eqno(56)$$

Consider now the vertex insertion graphs, Fig.~13.  Using (2) and
(15), the
amplitude from the box diagram contributing to the color factor term
$Tr(abcd)Tr({\bf 1})$ is $${\cal M}^V={1\over 16\pi^2}\int D^3 \alpha
{S^V(J,p)\over \alpha _1 \alpha _2s^2},\eqno(57)$$ with the numerator
given by
$S^V(J,p)=Tr(abcd)Tr({\bf 1})g^4V$, and $V$ to be the spin factor
contribution
from all the vertex insertion diagrams.  Using the spin factors for
the
color-oriented vertices in Fig.~2, and the expression for the
currents in
eq.~(15), one gets $$\eqalignno{ V_1&= -64\alpha_1\alpha_2 \alpha
_3,&\cr
V_2&=V_3= 16\alpha_3,&\cr V_4&= -16(\f \alpha _1+\alpha_2+\alpha
_3),&\cr V_5&=
16 \alpha _2(\f \alpha _1+\alpha _2+\alpha _3),&\cr V_6&=
V_9=V_{12}=V_{13}=
8\alpha_1 \alpha _2 \alpha _3,&\cr V_7&= -16(\alpha _1+\f \alpha
_2+\alpha_3),&\cr V_8&= 16\alpha_1(\alpha _1+\f \alpha _2+\alpha
_3),&\cr
V_{10}&= 16\alpha_1(\f\alpha _1+\alpha _2+\f\alpha _3),&\cr V_{11}&=
16\alpha_2(\alpha _1+\f\alpha _2+\f\alpha _3).&(58)\cr}$$ The sum is
$$V=\sum_{i=1}^{13}V_i=-32 \alpha _1 \alpha _2 \alpha _3.\eqno(59)$$

\def\g{{\cal G}}
\def\.{\cdot}
\def\bk#1{\langle #1\rangle}

The final result is $$\eqalignno{ {\cal M}^B&=Tr(abcd)Tr({\bf
1}){g^4\over
12\pi^2}s{[24]^2\over [12]\bk{23}\bk{34}[41]},&(60)\cr {\cal
M}^V&=Tr(abcd)Tr({\bf 1}){g^4\over 12\pi^2}t{[24]^2\over
[12]\bk{23}\bk{34}[41]},&(61)\cr {\cal M}&={\cal M}^B+{\cal
M}^V=Tr(abcd)Tr({\bf 1}){g^4\over 12\pi^2}(-u){[24]^2\over
[12]\bk{23}\bk{34}[41]}.&(62)\cr}$$

This result agrees with the result obtained by the string method
[19--21].  The
vanishing of the diagrams in Fig.~11 is also a feature shared by the
string
method.  In fact, it has been observed [21] that the string
expression for box
diagram corresponds to a calculation in the background Feynman gauge,
though a
mixture of the background gauge and the Neuveu-Gervais gauge seem to
be
required for the vertex-insertion diagram.  In the present case we
use the
background gauge throughout.  The total number of terms in the box
diagram (11)
and the vertex-insertion diagram (13) is quite comparible with that
using the
string method [20] ($2\times 14$ each) as well.  We conclude
therefore that in
most respects this method is just as efficient as the string method.

There is actually another similarity which is telling.  One notes
from (55) and
(56) that $B$ is proportional to $B_1$, so that many of the 11 terms
in (55)
combine to cancel one another.  The same happens in the
vertex-insertion
diagrams, and the same happens in the string approach.  This strongly
suggests
that as simple as the computation is, all together 24 non-vanishing
terms
rather than something of the order of $10^4$ which one has in the
ordinary
approach, there must be even simpler way of calculating things where
one can
avoid writing down terms that eventually cancel one another.  The
following
calculation shows how this can be attained.

We turn now to the computation of the box diagram with an internal
quark loop,
Fig.~14(a,b).  The result per flavor is obtained from (2) to be
$${\cal
M}^{'B}= {1\over 16\pi^2}\int D^4 \alpha {S^{'B}(J,p)\over (\alpha _2
\alpha
_4s+\alpha _1 \alpha _3t)^2},\eqno(63)$$ where $$\eqalignno{
S^{'B}(J,p)=-Tr(abcd)g^4\{&tr[(\gamma\epsilon_1)(\gamma J_1) (\gamma
\epsilon
_4)(\gamma J_4)(\gamma \epsilon _3)(\gamma J_3) (\gamma \epsilon
_2)(\gamma
J_2)]&\cr +&tr[(\gamma\epsilon_1)(\gamma J_2) (\gamma \epsilon
_2)(\gamma
J_3)(\gamma \epsilon _3)(\gamma J_4) (\gamma \epsilon _4)(\gamma
J_1)]\}.&(64)\cr}$$ Like (39), these traces are easily computable
using (14),
(38), and (49).  The first trace is $$\eqalignno{
&tr[(\gamma\epsilon_1)(\gamma
J_1) (\gamma \epsilon _4)(\gamma J_4)(\gamma \epsilon _3)(\gamma J_3)
(\gamma
\epsilon _2)(\gamma J_2)]=&\cr
&-4(\alpha_1+\alpha_2)\alpha_3\alpha_4^2{
\bk{12}[24]\bk{14}[43]\bk{13}[32]\bk{13}[32]+
[23]\bk{31}[42]\bk{21}[34]\bk{41}[23]\bk{31}\over
[21]\bk{12}\bk{13}\bk{14}}=&\cr
&8(\alpha_1+\alpha_2)\alpha_3\alpha_4^2s^2t{[24]^2\over
[12]\bk{23}\bk{34}[41]}.  &(65)}$$ The second trace is equal to the
first
trace.  Therefore, $${\cal M}^{'B}=-2{\cal M}^B/Tr({\bf
1}).\eqno(66)$$
Similarly, one can compute Fig.~14(c,d) to get $${\cal
M}^{'V}=-2{\cal
M}^V/Tr({\bf 1}).\eqno(67)$$ The equalities in (66) and (67) are easy
to
understand by using supersymmetry arguments[12,16,17,25].  Quarks and
gluinos
have the same spacetime coupling with the gluon, though they carry
different
colors.  If we should replace the quark loop by a gluino loop, the
only change
after we replace the color factor $Tr(abcd)$ in the quark loop by the
color
factor $Tr(abcd)Tr({\bf 1})$ in the gluino loop would be an extra
factor of
1/2, reflecting the majorana nature of the gluino.  Since the
four-gluon
amplitude in a pure supersymmetric QCD theory (gluinos are present
but not
quarks) is zero, the gluon/ghost loop contribution is equal and
opposite to the
gluino loop contribution.  Putting these two facts together, the
equality of
(66) and (67) are obtained.

These arguments can be reversed and to be used to compute ${\cal
M}^B$ and
${\cal M}^V$ from ${\cal M}^{'B}$ and ${\cal M}^{'V}$.  This
simplifies the
calculation of the pure gluon amplitudes because the quark loop box
diagram
contains only four terms, which are equal, instead of the 11 terms in
Fig.~12.
This also explains why many of these terms in (55) (similarly (58))
add up to
give zero.

\bigskip\bigskip
\centerline{\bf 6. Conclusions}
\bigskip
For high energy scatterings lepton and light-quark masses can be
ignored.
Chirality is then conserved and tremendous simplifications in the
calculations
of these amplitudes can be obtained by the use of the spinor-helicity
techniques [1--21].  With one exception [19--21], this technique was
used only
to calculate the tree amplitudes [1--18], because loop graphs contain
off-shell
momenta where chirality is not conserved and this technique cannot be
applied.
The exception [19--21] makes use of the string theory and is
applicable to
certain one-loop processes.  We have developed in this paper a
technique,
making use of the Feynman-parameter representation of a scattering
amplitude to
avoid the off-shell internal momental, to enable to spinor-helicity
method to
be used for {\it any Feynman diagram with any number of loops}.

Graphical methods
are used throughout to organize the terms and to avoid treading into
the
algebraic tangle.  The method was applied to a one-loop four-gluon
amplitude to
show that the present method is at least as efficient as the string
technique.
Application of the method to the calculation of other processes is
underway.

\bigskip\bigskip
\centerline{\bf Acknowledgements}
\bigskip
I am grateful to Z. Bern, K. Dienes,

D. Kosower, and T.-M. Yan for useful discussions.
This work is supported in part by the Natural Sciences and
Engineering
Research Council of Canada and the Qu\'ebec Department of Education.
Part of this manuscript was prepared during my visit to the Institut
des Hautes \'Etudes Scientifiques, Bures-sur-Yvette, France.

I wish to thank Profs. L. Michel and
M. Berger for their kind hospitality.

\vfill\eject
\centerline{\bf References}
\bigskip
\def\i#1{\item{[#1]}}
\def\npb#1{{\it Nucl.~Phys. }{\bf B#1}}
\def\plb#1{{\it Phys.~Lett. }{\bf #1B}}
\def\prl#1{{\it Phys.~Rev.~Lett. }{\bf B#1}}
\def\prd#1{{\it Phys.~Rev. }{\bf D#1}}
\def\pr#1{{\it Phys.~Rep. }{\bf #1}}
\def\ibid#1{{\it ibid.} {\bf #1}}

\i 1 F.A. Berends, R. Kleiss, P. De Causmaecker, R. Gastmans,
W. Troost, and T.T. Wu, \plb{103} (1981), 124;
\npb{206} (1982), 61;
\ibid{239} (1984), 382; \ibid{239} (1984), 395; \ibid{264} (1986),
243;
\ibid{264} (1986), 265.
\i 2  P. De Causmaecker, R. Gastmans,
W. Troost, and T.T. Wu, \plb{105} (1981), 215; \npb{206} (1982), 53.
\i{3} Z. Xu, D.-H. Zhang, and L. Chang, Tsinghua University
Preprints,
Beijing, China, TUTP-84/4, TUTP-84/5, TUTP-84/6; \npb{291} (1987),
392.
\i 4 F.A. Berends and W.T. Giele, \npb{294} (1987), 700; \ibid{306}
(1988),
759; \ibid{313} (1989), 595.
\i 5 F.A. Berends, W.T. Giele, and H. Kuijf, \plb{211} (1988), 91;
\ibid{232}
(1989), 266; \npb{321}
(1989), 39;\ibid{333} (1990), 120.
\i 6 J. Gunion and J. Kalinowski, \prd{34} (1986), 2119.
\i {7} J. Gunion and Z. Kunszt, \plb{159} (1985), 167; \ibid{161}
(1985), 333;
\ibid{176} (1986) 477.
\i{8} K. Hagiwara and D. Zeppenfeld, \npb{313} (1989), 560.
\i{9} R. Kleiss and H. Kuijf, \npb{312} (1989), 616.
\i{10} R. Kleiss and W.J. Stirling, \npb{262} (1985), 235.
\i{11} D. Kosower, \npb{315} (1989), 391; \ibid{335} (1990), 23.
\i{12} Z. Kunzt, \npb{271} (1986), 333.
\i{13} M. Mangano, \npb{309} (1988), 461.
\i{14} M. Mangano, S. Parke, and Z. Xu, \npb{298} (1988), 653.
\i{15} M. Mangano and S.J. Parke, \npb{299} (1988), 673; \prd{39}
(1989), 758.
\i{16} M. Mangano and S.J. Parke, \pr{200} (1991), 301.
\i{17} S. Parke and T. Taylor, \plb{157} (1985), 81; \npb{269}
(1986), 410;
\prl{56} (1986), 2459; \prd{35} (1987), 313.
\i {18} C. Dunn and T.-M. Yan, \npb{352} (1989), 402.
\i {19} Z. Bern and D.K. Kosower, \prl{66} (1991), 1669; preprint
Fermilab-Pub-91/111-T.
\i {20} Z. Bern and D.K. Kosower, preprint Fermilab-Conf-91/71-T.
\i {21} Z. Bern and D.C. Dunbar, preprint Pitt-91-17.
\i{22} C.S. Lam and J.P. Lebrun, {\it Nuovo Cimento} {\bf 59A}
(1969), 397.
\i{23} L.F. Abbott, \npb{185} (1981), 189.
\i{24} J. Paton and Chan Hong-Mo, \npb{10} (1969), 519.
\i{25} M.T. Grisaru and H.N. Pendleton, \npb{124} (1977), 81.

\vfill\eject

\bigskip\bigskip
\centerline{\bf Figure Captions}
\bigskip
\i{Fig.~1} Diagrams used to illustrate the Feynman-parameter rules.
(a) a box diagram; (b) a vertex diagram.
\i{Fig.~2} Vertices for QCD in the background Feynman gauge.
A thin solid line represents a gluon, a thick solid line represents
a quark, and a dashed line represents a Fadeev-Popov ghost. Gluon
lines with a circled `A' are external lines; those in the same

diagram without a circled `A' are internal lines. Each gluon line
in a diagram without any circled `A's present can be taken either as
an
external or an internal line, provided such a combination
of external and internal lines has not appeared
already in diagrams where explicit circled `A's appear.
\i{Fig.~3} Color-oriented vertices and their spin factors. A
line continuing through the vertex represents a $g_{\rho \sigma }$
factor; a line terminated at a heavy dot at the vertex represents
a vector written below the diagram. Other numerical factors for the
vertex also appear below the diagrams. The line labelled `1' carries
a
momentum $p_1$, a spacetime index $\alpha $, and a color index a.
Similarly, a line labelled `2' carries a momentum $p_2$, a spacetime
index $\beta $, and a color index b, etc.

For example, the spin factor for diagram (a) is
 $(p_2-p_3)_\alpha g_{\beta\gamma}$; the spin factor for diagram (e)
 is $+2g_{\alpha \gamma }g_{\beta \delta }$.
\i{Fig.~4} The color factor ${\cal C}$ for a gluon tree diagram.
\i{Fig.~5} A gluon tree diagram and its color factor used to
illustrate the proof of eq.~(22).
\i{Fig.~6} The color factor ${\cal C}$ for a tree diagram containing
a number of gluon trees attached to a quark line.
\i{Fig.~7} The $U(N)$ color factor ${\cal C}$ for a tree diagrams
with
two quark lines.
\i{Fig.~8} The color factor ${\cal C}$ for a one-loop diagram with
a quark line.
\i{Fig.~9} The color factor ${\cal C}$ for a one-loop diagram without
a quark line.
\i{Fig.~10} One-gluon-loop diagrams and the vanishing spin-flow
paths A,B,C,D.
Background Feynman gauge is used; the helicity configuarations
as well as e e choice of the reference momenta are discussed in the
text.
\i{Fig.~11} These diagrams make no contributions to the process
calculated
in Sec.~5.
\i{Fig.~12} Non-vanishing spin-flow paths (diagrams 1 to 11)
 for the one-gluon-loop box diagram
calculated in Sec.~5.
\i{Fig.~13} Non-vanishing spin-flow paths (diagrams 1 to 13)
for the one-gluon-loop vertex-insertion diagram calculated in Sec,~5.
\i{Fig.~14} One-fermion-loop diagrams for the processes calculated
in Sec.~5.

\end